\documentclass[12pt]{iopart}

\usepackage{citesort}

\usepackage{graphicx}

\usepackage{iopams}

\begin{document}
\title[Coexistence of LRMO and SC from Campbell length]{Coexistence of Long-Range Magnetic Order and Superconductivity from Campbell Penetration Depth Measurements\footnote{On the occasion of his 80th birthday, we dedicate this paper to Alexey Alexeevich Abrikosov who discovered two types of superconductors and described vortices in type-II superconductors \cite{Abrikosov57English,Abrikosov57Russian}.}}

\author{R Prozorov, M D Vannette, R T Gordon, C Martin, S L Bud'ko, P C Canfield}

\address{Ames Laboratory and Department of Physics \& Astronomy, Iowa State University, Ames, IA 50011, U.S.A}

\ead{prozorov@ameslab.gov}

\begin{abstract}
Application of a tunnel-diode resonator (TDR) technique for studies of the vortex response in magnetic superconductors is described. Operating at very small excitation fields and sufficiently high frequency, TDR was used to probe small-amplitude linear AC response in several types of single crystals where long-range magnetic order coexists with bulk superconductivity. Full local - moment ferromagnetism destroys superconductivity and can coexist with it only in a narrow temperature range ($\sim 0.3$ K). In contrast, weak ferromagnetic as well as antiferromagnetic orders can coexist with bulk superconductivity and may even lead to enhancements of vortex pinning. By analyzing the Campbell penetration depth we find sharp increase of the true critical current in the vicinity of the magnetic phase transitions. We conclude that critical magnetic fluctuations are responsible for this enhancement.
\newline\newline
\textit{Date: 12 September 2008}
\end{abstract}


\submitto{\SUST - ICSM-2008 Special issue.}

\maketitle

\section{Introduction}
Coexistence of bulk superconductivity (SC) and long-range magnetic order (LRMO) was studied by many researchers over past half century \cite{Ginzburg1957,Anderson1959,Bulaevskii1985,Sinha1989,Fischer1990,Maple1995,Lynn1997,Canfield1998,Muller2001,Kulic2006,prozorov2008}. In fact, this topic has been the subject of so many works and in so many materials that we have to apologize beforehand for inadvertent omission of some key references. While full local-moment ferromagnetism (LMFM) can coexist with superconductivity only in narrow temperature and field intervals, antiferromagnetic (AFM) and weak and/or itinerant ferromagnetic (IFM) order can occupy significant portions of the $H-T$ phase diagram in many superconductors. Magnetic superconductors can be classified according to their transition temperatures. Let us use $T_{SC}$ for the superconducting transition, $T_C$ for the Curie temperature of a ferromagnet and $T_N$ for the N\`{e}el temperature of an antiferromagnet. Then antiferromagnetic superconductors are the materials with $T_N < T_{sc}$ (e.g., ErNi$_2$B$_2$C, $T_N$= 6 K and $T_{SC}$= 11 K), superconducting antiferromagnets with $T_N > T_{sc}$ (e.g., DyNi$_2$B$_2$C, $T_N$= 12 K and $T_{SC}$= 6 K), superconducting ferromagnets with $T_C > T_{sc}$ (e.g., Y$_9$Co$_7$, $T_C$= 8 K and $T_{SC}$= 3 K) and ferromagnetic superconductors with $T_C < T_{sc}$ (e.g., ErRh$_4$B$_4$, $T_C$= 1.1 K and $T_{SC}$= 8.5 K). Full local-moment ferromagnetic superconductors are rare. In addition to ErRh$_4$B$_4$ \cite{Fertig1977}, there is Ho$_{x}$Mo$_{6}$S$_{8}$ \cite{Ishikawa1977} ($T_C\approx0.7$ K, $T_{SC}\approx 1.8$ K). Other types of coexisting phases are more abundant with borocarbides being among most interesting due to their robust ambient-pressure and relatively high $T_{SC}$ superconductivity, weak pinning and availability in (clean) single crystal form.

Practically all techniques used in low-temperature solid-state physics were employed to study magnetic superconductors (see, for example, \cite{Bulaevskii1985,Sinha1989,Fischer1990}). The general consensus is that local-moment ferromagnetism destroys superconductivity (at least with singlet pairing) due to spin-flip pair breaking, so the coexisting region is narrow, but finite \cite{Bulaevskii1982,Bulaevskii1985}. Close to the FM boundary various exotic effects are possible. For example, Fulde-Ferrell-Larkin-Ovchinnikov (FFLO) spatially - modulated superconductivity with finite pairing momentum, a crossover from type-II to type-I superconductivity as well as unusual spin configurations of the ferromagnetic subsystem \cite{Bulaevskii1985,Sinha1989,Fischer1990}. Weak and/or itinerant ferromagnetic state can coexist with superconductivity more easily also developing spatially inhomogeneous configurations (spiral or domain-like) on the length scales less than the superconducting coherence length, so that the effect of the exchange field on Cooper pairs is reduced. AFM order can coexist with superconductivity in an even wider range of materials. Of course, the spin configurations may be quite different from simple parallel or antiparallel alignment and various metamagnetic transitions can still be found in the superconducting region of the H-T diagram. Adding the effect of anisotropy, both superconducting and magnetic, leads to a variety of interesting effects and coexisting phases.

In this paper we describe sensitive tunnel-diode resonator technique applied to study small-amplitude linear AC response in the vortex state of magnetic superconductors. It directly probes dynamic magnetic susceptibility which, in type-II superconductors, is determined by the vibrations of Abrikosov vortices in their potential wells, usually expressed in terms of the so-called Campbell penetration depth \cite{blatter94,Brandt1995}. In the absence of vortices, the same measurement probes the London penetration depth. We will describe results obtained on several magnetic superconductors and show that such measurements can not only be used for precision mapping of the $H-T$ phase diagram, but also serve to study the mutual influence of LRMO and SC deep in the superconducting state.

\section{Campbell penetration depth}

The response of an elastic medium in the presence of disorder is a general problem in physics, applicable to charge density waves, dislocations, ferromagnetic domain walls and vortices in superconductors. In unconventional and magnetic  superconductors, strong thermal, magnetic and quantum fluctuations add new levels of complexity \cite{blatter94,Brandt1995,Yeshurun1996}. Usually linear elastic response in sufficient to explain the data. In particular, low-amplitude AC response assumes the validity of Hooke's law \cite{Labusch1968,Campbell1969,Campbell1971,Campbell1972,Koshelev1991,Brandt1995,Prozorov2003}, so the reaction of Abrikosov vortices to small perturbations is perfectly elastic. Vortices transmit these perturbation caused by a small $AC$ field at the surface as either compressional or tilt waves (or both), depending upon the geometry of the experiment \cite{Brandt1995,Brandt1991}. In both cases the AC response has been calculated by several authors \cite{blatter94,Brandt1995,Brandt1991,Coffey1991,Coffey1992,Koshelev1991,Prozorov2003} and is given to a good approximation by $\lambda^2 = \lambda_L^2 +\lambda_{vortex}^2$, where $\lambda_L (T)$ is the London penetration depth and $\lambda_{vortex}^2$ is the extent of vortex - transmitted perturbation, given by \cite{Coffey1991,Coffey1992}
\begin{equation}
\lambda_{vortex}^2 = \frac{B^2}{4\pi \alpha}
 \frac{1 - i \omega/\omega_{pin} \exp{(-U/k_BT)}}{1 + i \omega /\omega_{pin}}\\
\end{equation}
\noindent where pinning is parameterized by the Labusch constant per unit volume $\alpha$ \cite{Labusch1968}, $\omega_{pin} =\alpha/\eta$ is the pinning frequency and $\eta$ is the viscous drag coefficient. $\omega_{pin}$ is typically $10^9$ Hz or higher \cite{Owliaei1992}. $U(T,B)$ is the vortex activation energy that determines the rate of thermal activation. This term becomes important near the usual irreversibility line and gives rise to a large increase in $\lambda_{vortex}$. The effects discussed in this paper are roughly 1000 times smaller and occur at temperatures well below $T_{SC}$ where $U(T,B)$ is large and $\omega_{pin} \exp{(-U/k_{B}T)} \ll \omega \ll \omega_{pin}$ where our working frequency $\omega /2\pi \approx 10$ MHz. The vortex response is then dominated by the Campbell length \cite{Brandt1995,Brandt1991,Campbell1971,Campbell1969,Koshelev1991}, $\lambda^2_C = C_{xx}/\alpha$ where $C_{xx}$ is the relevant elastic module, $C_{11}$ for compression (excitation field is parallel to the vortices) or $C_{44}$ - tilt module (AC field is perpendicular to the vortices). Both moduli are approximately equal to $B^2/4\pi$ and with the Labusch parameter, $\alpha=Bj_c/cr_p$\cite{Labusch1968}, we obtain a widely observed $\sqrt{B}$ dependence to the penetration depth in the vortex pinning regime. The radius of the pinning potential, $r_p$ is usually taken to be the size of the vortex core or the coherence length, $\xi$. The true critical current density, $j_c$, as opposed to that estimated from $M(H,T)$ relaxed persistent current density, can thus be obtained from the measurements of $\lambda_C$. This approach can be further generalized by taking into account the non-parabolic shape of the pinning potential and presence of the Bean biasing current in a non-uniform vortex distribution \cite{Prozorov2003}. It should be noted that certain sharp features observed in measured $\lambda(T,B)$ can be interpreted as abrupt (or even discontinuous) changes in the elastic moduli related, for example, to the transformation from triangular to square flux lattice \cite{Prozorov2007a}.

For the purpose of this paper, we will therefore use the expression applicable deep in the superconducting state, far from flux flow regime, because then $\lambda_L \ll \lambda_C$.

\begin{equation}
\lambda^2 = \lambda_L^2+\lambda_C^2 \approx \xi B\frac{c}{4\pi j_c}
\label{eq:lambda}
\end{equation}

\section{Experimental}
\subsection{Tunnel-diode Resonator Technique}
Radio frequency dynamic magnetic susceptibility, $\chi $, was measured by using sensitive tunnel diode resonator (TDR) technique. The design and capabilities of TDR are discussed elsewhere \cite{vandegrift,Prozorov2000,Prozorov2000a,Prozorov2006a}. In brief, the resonance is maintained by a tunnel diode that exhibits negative differential resistance when properly biased, and thus acts as a low current AC power source that compensates for losses in an $LC$ tank circuit. As a result, the circuit self-resonates at the natural resonant frequency, $2\pi f_0=1/\sqrt{LC}$ and the excitation field of the inductor, $L$, is very low, $\sim 20$ mOe. This small excitation field is especially advantageous when studying vortices in superconductors, because it is not strong enough to displace them out of the potential wells, so it is only probing vortex oscillations about their static positions. This is known as the Campbell regime. Conventional AC techniques driven by external power sources usually require relatively large excitation fields (0.1-10 Oe), because they rely on the measurements of the amplitude, whereas the TDR technique is based on the measurements of the frequency shift. A properly designed and stabilized circuit allows one to measure changes in dynamic magnetic susceptibility on the order of a few parts per billion. For typical crystals ($\sim 1$ mm in size) this translates into sub-Angstr\"om resolution of the penetration depth or pico-emu sensitivity to the changes in the magnetic moment.

The sample to be studied is mounted on a sapphire rod with a small amount of
low temperature grease and inserted into a small copper coil which acts as
the inductor in the $LC$ tank circuit. Changes in the magnetic susceptibility of the sample induce changes in the resonant frequency of the $LC$ circuit. It is straightforward to show that,

\begin{equation}
\frac{\Delta f}{f_{0}}\approx -1\frac{V_{s}}{2V_{c}\left(1-N\right)}4\pi \chi .
\end{equation}%
\noindent where $\Delta f=f\left( H,T\right) -f_0$ is the change in the resonant
frequency due to the sample, $f_0$ is the resonant frequency of an empty coil, $V_s$ is the volume of the sample, $V_c$ is the volume of the coil and $N$ is the demagnetization factor. Magnetic susceptibility of a superconductor in the limit of small excitation field (linear AC response) can be written as,

\begin{equation}
4\pi \chi  \simeq \mu \frac{{\lambda \left( \mu  \right)}}{R}\tanh \frac{R}{{\lambda \left( \mu  \right)}} - 1 .
\end{equation}%
\noindent where $\mu$ is the normal-state magnetic permeability of the material (which can be relevant for magnetic superconductors) and $\lambda$ is the AC penetration depth. The effective dimension $R$ takes into account penetration of the magnetic field not only from the sides, but also from top and bottom surfaces in a finite sample \cite{Prozorov2000}. In the linear response, $\lambda(\mu)=\lambda/\sqrt{\mu}$, where $\lambda$ is the AC penetration depth of a non-magnetic sample\cite{Bulaevskii1985}. Therefore, by measuring the frequency shift, we can directly probe the AC penetration depth.

\subsection{Samples}

All samples used in this study were grown in US DOE Ames Laboratory. Borocarbide single crystals ($RNi_2B_2C$, $R=Er,Tm$) were grown out of Ni$_{2}$B flux \cite{Cho1995}. Detailed discussions of the superconducting properties with the emphasis on the interplay between superconductivity and magnetism as well as comparison to nonmagnetic borocarbides can be found in Refs.~\cite{Lynn1997,Canfield1998,Muller2001,Bud'ko2006}. Single crystals of ErRh$_4$B$_4$ were grown at high temperatures from a molten copper flux as described in Refs.~\cite{Okada1996,Shishido1997}.

\section{Results and Discussion}

We now show results obtained on some magnetic superconductors. In a typical experiment the sample is cooled in zero applied magnetic field and then an external DC magnetic field is applied and kept constant throughout warming and cooling when the data are collected (known as zfc-fc process). The warming up and cooling down cycle may be repeated several times to study possible hysteretic behavior.

\subsection{ErNi$_2$B$_2$C}

We begin with ErNi$_2$B$_2$C, which exhibits transition to an antiferromagnetic state with spins along the crystallographic $b-$axis at $T_N \approx$ 6 K, deep inside the superconducting phase that appears at $T_{SC} \approx$ 11 K. At lower temperatures, below $T_C \approx$ 2.2 K, a weak ferromagnetic order appears. Thus ErNi$_2$B$_2$C can be classified as both antiferromagnetic and weak ferromagnetic superconductor. The existence of both LRMO phases were directly detected by Bitter decoration \cite{Veschunov2007}, neutron diffraction \cite{Lynn1997,Choi2001} and Hall probe studies \cite{Bluhm2006}. A detailed $H-T$ phase diagram shows significant impact of the long-range magnetic order on the anisotropic superconducting properties \cite{Bud'ko2006}.

\begin{figure}[tb]
\begin{center}
\includegraphics[width=16cm]{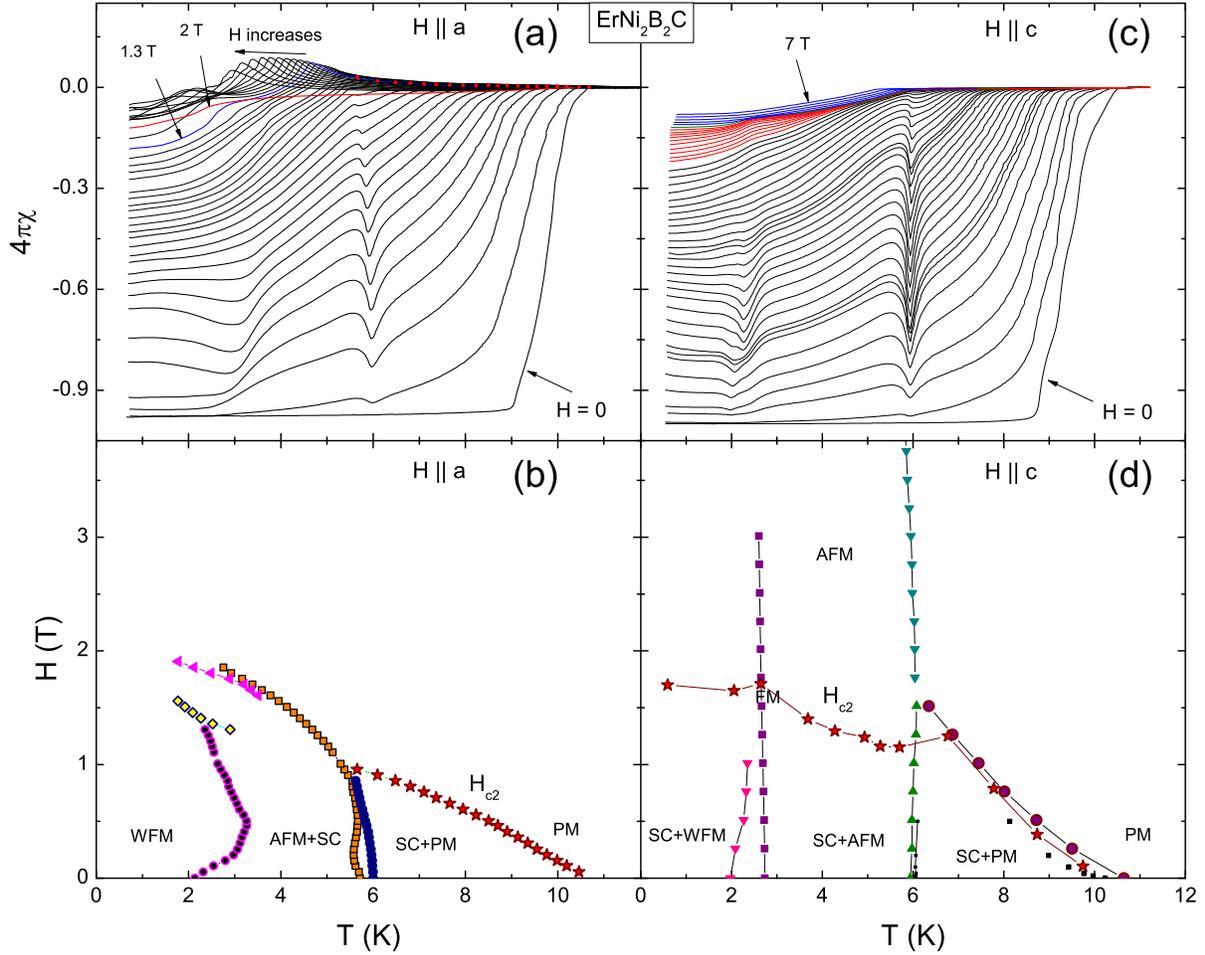}
\end{center}
\caption{(a) Dynamic magnetic susceptibility, $4 \pi \chi(T)$, measured in different applied DC fields in a ErNi$_2$B$_2$C single crystal. External DC and AC magnetic fields were applied along the crystallographic $a-$axis. (b) $H-T$ phase diagram constructed by mapping various features detected in (a). (c) similar to (a), but magnetic field is applied along the $c-$axis. (d) $H-T$ phase diagram constructed from (c).}
\label{ernibc}
\end{figure}

Figure \ref{ernibc} summarizes measurements of the dynamic magnetic susceptibility, $4 \pi \chi$, as well as $H-T$ diagrams constructed from these measurements. Two prominent features can be see in $4 \pi \chi (T)$ in the presence of vortices, whereas nothing appears in $H=0$ curves for both orientations. We, therefore, do not find any evidence for spontaneously generated vortices at either W-FM or AFM transitions. This is consistent with miniature Hall-probe studies \cite{Bluhm2006} as well as measurements of the surface impedance at microwave frequencies \cite{Jacobs1995}. A prominent dip in the response at about 6 K evidently marks the antiferromagnetic transition. While low-field decoration detected accumulation of vortices in the ordered phase along the AFM twin boundaries, which was interpreted as the enhancement of pinning, our results suggest that this pinning is either weak, significantly field dependent, or the density of such pinning centers is insufficient to result in a macroscopic enhancement of the critical current. (If bulk pinning were to develop below $T_N$, we would observe a step-like decrease in the penetration depth, see Eq.~\ref{eq:lambda}). However, we only see the effect in the immediate vicinity of the magnetic phase transition. 

We propose that this reduction of the Campbell length at $T_N$ is caused by the enhancement of pinning due to large magnetic fluctuations accompanying this second-order transition. In a collective pinning theory, pinning comes from the mean square variation in the distribution of the normal pinning centers with concentration $n_i$ and leads to $j_c  =  j_0 \left( \xi/L_c \right)^2  \sim \left( n_i \right)^{(2/3)}$ \cite{blatter94}. Here $j_0$ is the depairing current density and $L_c$ is the collective pinning length. In the vicinity of the LRMO phase transition, in addition to the condensation energy, there is an additional magnetic part of the pinning. Detailed description of this mechanism of magnetic fluctuations - mediated enhancement of the pinning strength will be reported elsewhere \cite{Prozorov2008a}. On the other hand, below the weak ferromagnetic transition, low-field data show a step-like feature that is consistent with the development of the bulk pinning. It was  also demonstrated in Bitter decoration experiments \cite{Veschunov2007} as well as transport and magnetization measurements \cite{Gammel2000}. It is worth noting that in the case of an external field applied along the magnetic easy axis, Fig.~\ref{ernibc}a, positive signal develops at high fields, probably due to metamagnetic transition in the spin structure.

\subsection{TmNi$_2$B$_2$C}

We now discuss the results obtained in TmNi$_2$B$_2$C single crystals. In this antiferromagnetic superconductor, $T_{SC}\approx$ 11 K and $T_N\approx$ 1.8 K. Detailed neutron diffraction studies show that contrary to ErNi$_2$B$_2$C crystals, in  TmNi$_2$B$_2$C spins orders along the crystallographic $c-$axis \cite{Lynn1997}. Perhaps it is this fact that results in a rich low-temperature magnetic phase diagram.  The magnetic phase diagram has been reported in several studies, for example in Ref.\ \cite{Eskildsen1998}

\begin{figure}[tb]
\begin{center}
\includegraphics[width=16cm]{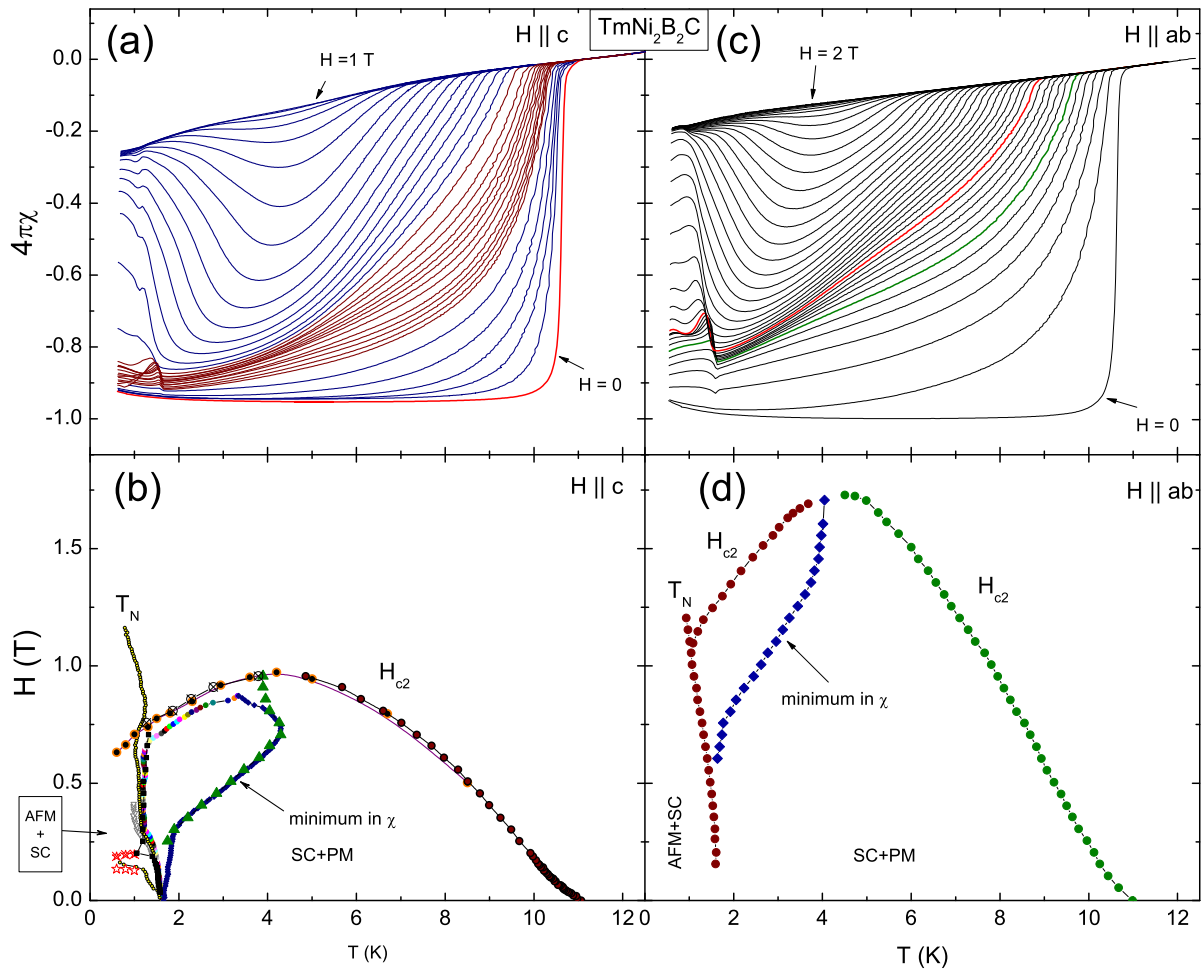}
\end{center}
\caption{(a) $4 \pi \chi(T)$, measured in different applied DC fields in a TmNi$_2$B$_2$C single crystal. External field was applied along the crystallographic $c-$axis. (b) $H-T$ phase diagram constructed by mapping various features detected in (a). (c) similar to (a), but magnetic field is applied along the $ab-$plane. (d) $H-T$ phase diagram constructed from (c).}
\label{tmnibc}
\end{figure}

Figure \ref{tmnibc} summarizes our measurements in a way similar to Fig.~\ref{ernibc} showing the results for a magnetic field applied along the magnetic easy $c-$axis in panels (a) and (b) and in the perpendicular orientation, panels (c) and (d). Similarly to ErNi$_2$B$_2$C, the transition to the ordered phase is marked by the decrease in the penetration depth around $T_N$. However, at the lower temperatures more structure appears, especially at the lower fields. The situation is complicated by the possible existence of two different magnetic moments as detected by inelastic neutron scattering and muon spin relaxation techniques \cite{Gasser1998}.

\subsection{ErRh$_4$B$_4$}

Finally we show the results obtained in a full local-moment single crystals of a ferromagnetic superconductor ErRh$_4$B$_4$. Detailed investigation of the narrow coexisting region between FM and SC phased by using the tunnel-diode resonator is reported elsewhere \cite{prozorov2008}. Here we focus on a comparison of this ferromagnetic superconductor in the entire temperature range with the discussed above magnetic borocarbides.

\begin{figure}[tb]
\begin{center}
\includegraphics[width=16cm]{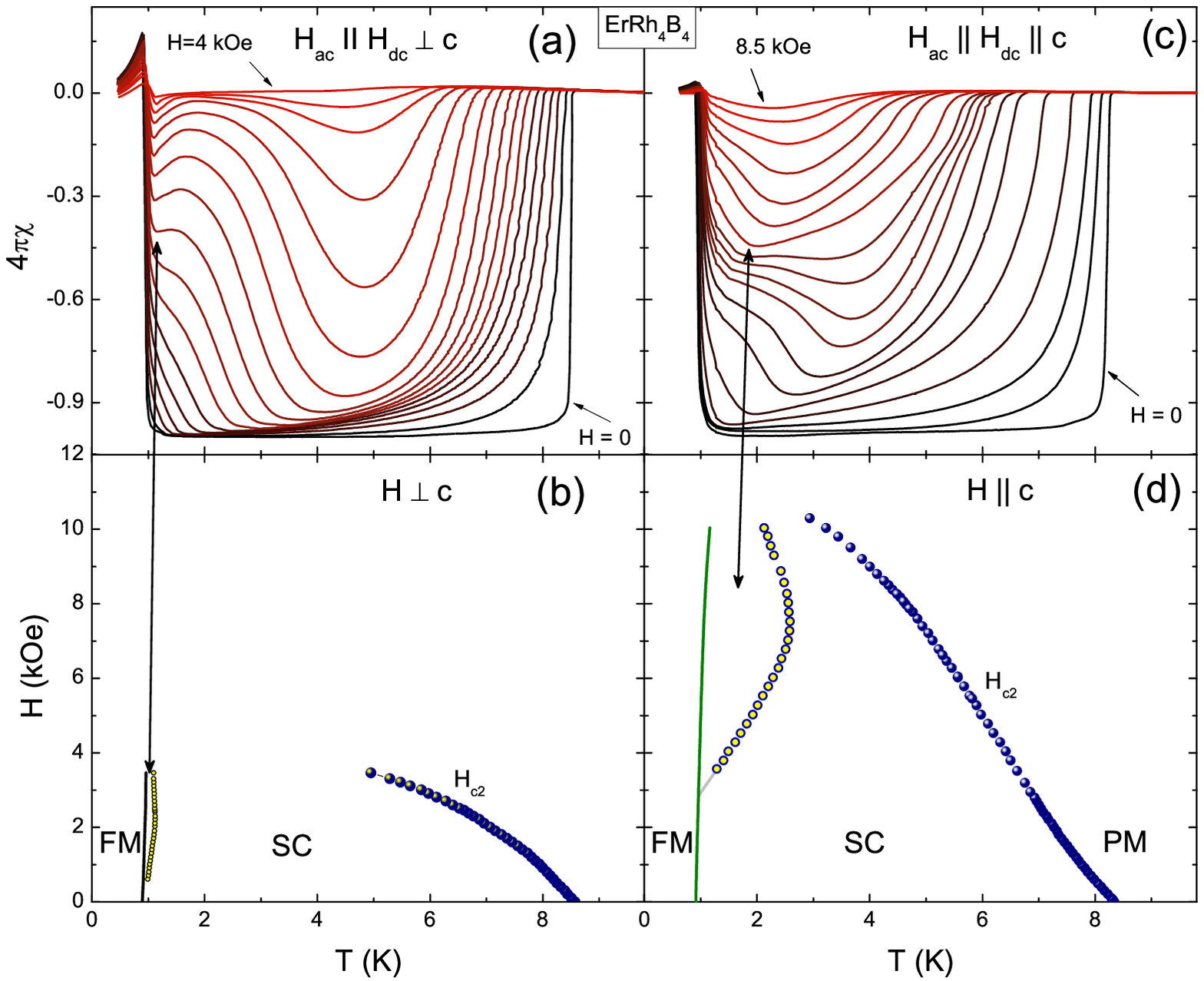}
\end{center}
\caption{(a) $4 \pi \chi(T)$, measured in different applied DC fields in a ErRh$_4$B$_4$ single crystal. External field was applied along the crystallographic $a-$axis (magnetic ordering axis). (b) $H-T$ phase diagram constructed by mapping various features detected in (a). (c) similar to (a), but magnetic field is applied along the $c-$axis. (d) $H-T$ phase diagram constructed from (c).}
\label{errhb}
\end{figure}

ErRh$_4$B$_4$ becomes superconducting at $T_{SC}$= 8.5 K and undergoes ferromagnetic transition at about $T_C$= 1 K, which apparently destroys superconductivity. Er$^{3+}$ ions carry a full local magnetic moment of $8 \mu_B$, almost equal to the free ion moment of $9 \mu_B$. The ferromagnetic easy axis is the crystallographic $a-$axis. Figure \ref{errhb} provides information similar to the previous figures allowing for easy comparison.

First of note is the transition to a ferromagnetic state that also shows an increasing magnetic susceptibility approaching $T_C$. This is a typical feature of the TDR measurement performed on local-moment ferromagnets \cite{Vannette2008}. Comparing Fig.~\ref{errhb}(a) and (c), one can see that due to magnetic anisotropy the degree of this paramagnetic enhancement depends on the orientation of the magnetic field with respect to a magnetic easy axis. While narrow coexisting region exhibits very interesting behavior, such as significant asymmetry and hysteresis between warming and cooling through $T_C$ and possible transition to a type-I superconducting state \cite{prozorov2008}, here we focus on the behavior deeper inside the superconducting phase. A broad minimum in $4 \pi \chi (T)$ around 5 K is simply due to a competition between the simultaneous increase of both critical current and paramagnetic magnetic permeability on cooling.  The latter ultimately wins at $T_C$. However, just before it happens, there is another minimum in the susceptibility, most prominent in Fig.~\ref{errhb}(c) at the elevated fields. This behavior cannot be understood in terms of the critical current, although some possibility of pinning on ferromagnetic fluctuations above $T_C$, similar to previously discussed superconductors, still remains. An alternative explanation could be the development of an elusive FFLO state as predicted by Bulaevskii for this particular superconductor \cite{Bulaevskii1985}. If one compares the phase diagrams, Fig.~\ref{errhb}(b) and (d), where open circles mark the position of this second minimum to that published in Ref.~\cite{Bulaevskii1985} for ErRh$_4$B$_4$ for different demagnetization factors, there is an apparent similarity. Of course, this observation is only a hint requiring further detailed studies.

\section{Conclusions}
Comparing figures \ref{ernibc}, \ref{tmnibc} and \ref{errhb}, we conclude that precision measurements of the dynamic magnetic susceptibility imply that increasing out-of-$ab-$plane component of the rare-earth moment leads to a suppression of superconductivity. Most likely in compounds like TmNi$_2$B$_2$C some uncompensated moment develops at low temperatures and higher fields. However, pure antiferromagnetic transition seems to enhance the true critical current via additional magnetic pinning on critical fluctuations in the vicinity of $T_N$. This finding may provide some guidance to creating artificial AFM/SC structures which would operate around $T_N$ and in which critical current can be tuned to the desired value.

\ack
Discussions with John Clem, Vladimir Kogan, Kazushige Machida and Roman Mints are appreciated. Work at the Ames Laboratory was supported by the Department of Energy-Basic Energy Sciences under Contract No. DE-AC02-07CH11358. R. P. acknowledges support from the Alfred P. Sloan Foundation.


\section*{References}
\begin{thebibliography}{99}

\bibitem{Abrikosov57English}
Abrikosov A~A 1957 {\em Soviet Physics JETP (English)\/} {\bf 5} 1174-1182

\bibitem{Abrikosov57Russian}
Abrikosov A~A 1957 {\em Zh. Eksp. i Teor. Fiz\/} {\bf 32} 1442-1450

\bibitem{Ginzburg1957}
Ginzburg V~L 1957 {\em Soviet Phys. JETP\/} {\bf 4} 153

\bibitem{Anderson1959}
Anderson P~W and Suhl H 1959 {\em Phys. Rev.\/} {\bf 116} 898

\bibitem{Bulaevskii1985}
Bulaevskii L~N, Buzdin A~I, Kuli\`c M~L and Panjukov S~V 1985 {\em Adv. Phys.\/} {\bf 34} 175

\bibitem{Sinha1989}
Sinha K~P and Kakani S~L 1989 {\em Magnetic Superconductors: Recent
  Developments\/} (New York: Nova Science Publishers)

\bibitem{Fischer1990}
Fischer O 1990 {\em Magnetic superconductors\/} ({\em in "Ferromagnetic Materials"\/} vol 5, Ch. 6) (Amsterdam: Elsevier, 1990)

\bibitem{Maple1995}
Maple M~B 1995 {\em Physica B\/} {\bf 215} 110

\bibitem{Lynn1997}
Lynn J~W, Skanthakumar S, Huang Q, Sinha S~K, Hossain Z, Gupta L~C, Nagarajan R and Godart C 1997 {\em Phys. Rev. B\/} {\bf 55} 6584-6598

\bibitem{Canfield1998}
Canfield P~C, Gammel P~L and Bishop D~J 1998 {\em Physics Today\/} {\bf 51} 40

\bibitem{Muller2001}
Muller K~H and Narozhnyi V~N 2001 {\em Reports on Progress in Physics\/} {\bf
  64} 943--1008

\bibitem{Kulic2006}
Kulic M~L 2006 {\em C. R. Phys.\/} {\bf 7} 4

\bibitem{prozorov2008}
Prozorov R, Vannette M~D, Law S~A, Bud'ko S~L and Canfield P~C 2008 {\em Phys. Rev. B\/} {\bf 77} 100503

\bibitem{Fertig1977}
Fertig W~A, Johnston D~C, DeLong L~E, McCallum R~W, Maple M~B and Matthias B~T 1977 {\em Phys. Rev. Lett.\/} {\bf 38} 987

\bibitem{Ishikawa1977}
Ishikawa M and Fischer O 1977 {\em Solid State Commun.\/} {\bf 23} 37

\bibitem{Bulaevskii1982}
Bulaevskii L~N, Buzdin A~I, Panyukov S~V and Kulic M~L 1982 {\em Phys. Lett. A\/} {\bf 89A} 93

\bibitem{blatter94}
Blatter G, Feigelman M~V, Geshkenbein V~B, Larkin A~I and Vinokur V~M 1994 {\em Rev. Mod. Phys.\/} {\bf 66} 1125-1388

\bibitem{Brandt1995}
Brandt E~H 1995 {\em Rep. Prog. Phys.\/} {\bf 58} 1465-1594

\bibitem{Yeshurun1996}
Yeshurun Y, Malozemoff A~P and Shaulov A 1996 {\em Rev. Mod. Phys.\/} {\bf 68} 911

\bibitem{Labusch1968}
Labusch R 1968 {\em Phys. Rev.\/} {\bf 170} 470-474

\bibitem{Campbell1969}
Campbell A~M 1969 {\em J. Phys. C\/} {\bf 2} 1492-1501

\bibitem{Campbell1971}
Campbell A~M 1971 {\em J. Phys. C\/} {\bf 4} 3186-3198 

\bibitem{Campbell1972}
Campbell A~M and Evetts J~E 1972 {\em Critical currents in superconductors\/}, (Monographs on physics, Taylor \& Francis Ltd. (London))

\bibitem{Koshelev1991}
Koshelev A and Vinokur V 1991 {\em Physica C\/} {\bf 173} 465-475

\bibitem{Prozorov2003}
Prozorov R, Giannetta R, Kameda N, Tamegai T, Schlueter J and Fournier P 2003 {\em Phys. Rev. B\/} {\bf 67} 184501

\bibitem{Brandt1991}
Brandt E~H 1991 {\em Phys. Rev. Lett.\/} {\bf 67} 2219-2222

\bibitem{Coffey1991}
Coffey M~W and Clem J~R 1991 {\em Phys. Rev. Lett.\/} {\bf 67} 386

\bibitem{Coffey1992}
Coffey M~W and Clem J~R 1992 {\em Phys. Rev. B\/} {\bf 45} 10527-10535

\bibitem{Owliaei1992}
Owliaei J, Sridhar S and Talvacchio J 1992 {\em Phys. Rev. Lett.\/} {\bf 69} 3366-3369

\bibitem{Prozorov2007a}
Prozorov R, Kogan V, Vannette M, Bud'ko S and Canfield P 2007 {\em Phys. Rev. B.\/} {\bf 76} 094520

\bibitem{vandegrift}
VanDegrift C~T 1975 {\em Rev. Sci. Inst.\/} {\bf 48} 599-607

\bibitem{Prozorov2000}
Prozorov R, Giannetta R, Fournier P and Greene R 2000 {\em Phys Rev Lett\/} {\bf 85} 3700-3703 

\bibitem{Prozorov2000a}
Prozorov R, Giannetta R, Fournier P and Greene R 2000 {\em Physica C\/} {\bf 341-348} 1703-1704

\bibitem{Prozorov2006a}
Prozorov R and Giannetta R 2006 {\em Supercond Sci Technol\/} {\bf 19} R41

\bibitem{Cho1995}
Cho B~K, Canfield P~C, Miller L~L, Johnston D~C, Beyermann W~P and Yatskar A
  1995 {\em Phys. Rev. B\/} {\bf 52} 3684-3695

\bibitem{Bud'ko2006}
Bud'ko S~L and Canfield P~C 2006 {\em C. R. Phys.\/} {\bf 7} 56-67

\bibitem{Okada1996}
Okada S, Kudou K, Shishido T, Satao Y and Fukuda T 1996 {\em Jpns. J. Appl. Phys., Part 2\/} {\bf 35} L790

\bibitem{Shishido1997}
Shishido T, Ye J, Sasaki T, Note R, Obara K, Takahashi T, Matsumoto T and
  Fukuda T 1997 {\em J. Solid State Chem.\/} {\bf 133} 82

\bibitem{Veschunov2007}
Veschunov I~S, Vinnikov L~Y, Bud'ko S~L and Canfield P~C 2007 {\em Phys. Rev. B\/} {\bf 76} 174506 

\bibitem{Choi2001}
Choi S~M, Lynn J~W, Lopez D, Gammel P~L, Canfield P~C and Bud'ko S~L 2001 {\em Phys. Rev. Lett.\/} {\bf 87} 107001

\bibitem{Bluhm2006}
Bluhm H, Sebastian S~E, Guikema J~W, Fisher I~R and Moler K~A 2006 {\em Phys. Rev. B\/} {\bf 73} 014514

\bibitem{Jacobs1995}
Jacobs T, Willemsen B~A, Sridhar S, Nagarajan R, Gupta L~C, Hossain Z, Mazumdar
  C, Canfield P~C and Cho B~K 1995 {\em Phys. Rev. B\/} {\bf 52} R7022-R7025

\bibitem{Prozorov2008a}
Prozorov R, Vannette M~D, Mints R~G, Kogan V~G, Bud'ko S~L and Canfield P~C 2008 {\em to be published\/}

\bibitem{Gammel2000}
Gammel P~L, Barber B, Lopez D, Ramirez A~P, Bishop D~J, Bud'ko S~L and Canfield P~C 2000 {\em Phys. Rev. Lett.\/} {\bf 84} 2497-2500

\bibitem{Eskildsen1998}
Eskildsen M~R, Harada K, Gammel P~L, Abrahamsen A~B, Andersen N~H, Ernst G, Ramirez A~P, Bishop D~J, Mortensen K, Naugle D~G, Rathnayaka K~D~D and Canfield P~C 1998 {\em Nature (London)\/} {\bf 393} 242-245

\bibitem{Gasser1998}
Gasser U, Allenspach P and Furrer A 1998 {\em J. Alloys Compd.\/} {\bf 275-277} 587-590

\bibitem{Vannette2008}
Vannette M, Sefat A, Jia S, Law S, Lapertot G, Bud'ko S, Canfield P, Schmalian J and Prozorov R 2008 {\em J Magn Magn Mater\/} {\bf 320} 354-363

\end{thebibliography}
\end{document}